\documentstyle[12pt]{article}

\font\sqi=cmssq8
\def\DR{\rm I\kern-1.45pt\rm R}

\def\DC{\kern2pt {\hbox{\sqi I}}\kern-4.2pt\rm C}

\begin{document}
\begin{center}
\vspace{5cm}
{\Large On the Geometry of Relativistic Anyon}
\\
\vspace{3cm}
A.Nersessian{\footnote{Work is supported
by INTAS-RFBR grant No. 95-0829.}}
{\footnote{e-mail:nerses@thsun1.jinr.dubna.su}}\\
\vspace{1cm}
{\it Bogoliubov Laboratory of Theoretical Physics, JINR,  \\
Dubna, Moscow district, 141980 Russia}
                                           \\

\vspace{1cm}

\end{center}

\vspace{3cm}
\begin{center}
{Abstract}
\end{center}
A twistor model is proposed for the free relativistic anyon.
 The Hamiltonian reduction of this model by
the action of the spin generator leads to the minimal covariant model;
whereas that by the action of spin and mass generators, to
the anyon model with free phase space that is a cotangent bundle of
the Lobachevsky plane with twisted symplectic structure. Quantum
mechanics of that model is described by irreducible representations of
the  (2+1)-dimensional Poincare' group.
 \vspace{0.2cm}

\thispagestyle{empty}
\newpage
{\bf 1. Introduction.} As is known, in (2+1)-dimensional space-time there
can exist anyons, particles with arbitrary spins and statistics
\cite{anyon}. Quantum mechanics of anyons as a spinning particles
is described by unitary irreducible representations of the (2+1)-dimensional
Poincare' group \cite{jakiw}.
However, classical theory of anyons cannot be described within the
standard approach based on the introducing of Grassmann variables.
For this purpose are use the so-called "minimal" and "extended" approaches
(here me employ the classification of Ref.\cite{misha}).\\
In the minimal approach,  the spin is provided by the nontrivial
symplectic structure of phase space \cite{ss,jakiw}.
The minimal relativistic-covariant model (in what follows, $K-$model)
of the anyon with spin $s$  is realized on the phase space $T_*\DR^{1.2}$
with the twisted symplectic structure
\begin{equation}
g_{ab}dP^a\wedge dQ^b +
 s\frac{P^a}{P^3}\varepsilon_{abc}dP^b\wedge dP^c ,\quad
P=\sqrt{P^aP_a},\label{Rs}\end{equation}
where $g_{ab}=diag(1,-1,-1)$ is the metric of space $\DR^{1.2}$
parametrized by the anyon relativistic momentum $P^a$;
its position is defined by the fiber  coordinates $Q^a$; the generators
of Lorentz rotations
$J^{ab}=\frac{1}{2}\varepsilon^{abc}J_c$
are given by the relations
\begin{equation}
 J_a= \varepsilon_{abc}P^bQ^c +s\frac{P^a}{P},
\label{Kj}\end{equation}
and  the phase space is restricted by the mass constraint: $P^aP_a=m^2$.

In the extended approach  the spin is described by
the extra phase degrees of freedom and it value is fixed by an
appropriate constraints \cite{misha,plushchay,ch}.

A shortcoming of minimal models is their being nonlinear; whereas that of
extended models is the nontrivial structure of phase space and constraints.
As a result, even the description of the interaction of an anyon with
an external electromagnetic field in those approaches
is not a simple problem \cite{poly,ch}.

In this note, we propose a twstor model of a free relativistic
anyon that has no drawbacks of both extended and minimal models.
The phase space of this model (in what follows, $T-$model)
is the twistor space; whereas the constrains fixing the anyon
mass $m$ and spin $s$ and the (2+1)-dimensional Poincare'
group define a linear symplectic transformations.

The Hamiltonian reduction of the $T-$ model by  the action of
the spin generator results in the $K-$model both at the classical and
 quantum level.

The Hamiltonian reduction of the $T-$model by the action
of the spin and mass generators leads to the minimal model of an anyon
with a free phase space. Its phase space is a cotangent  bundle of the
Poincare' (or Klein) model of the Lobachevsky plane
${\cal L}=SU(1.1)/U(1)$ with the twisted symplectic
(pseudo-K\"ahler) structure and generators of the
Poincare' group define its isometries (${\cal L}-$model).
The anyon quantum mechanics in the ${\cal L}-$model is described
by the reducible representation of the  $SU(1.1)$ group on the
Lobachevsky plane playing the role of the  momentum space.
It can be obtained by the reduction of quantum-mechanical $T-$model.\\

 {\bf 2. $T-$model.} Consider the twistor space
${\bf T}=T\DC^{1.1}$ with the symplectic structure
 \begin{equation}
\Omega =i\eta_{\alpha{\bar\beta}}
(d\pi^\alpha\wedge d{\bar\omega}^{\bar\beta} +
 d{\omega}^{\alpha}\wedge d{\bar\pi}^{{\bar\beta}}), \quad \alpha,\beta,=0,1
 \label{1} \end{equation}
where $\eta-$ is the Pauli matrix $\sigma^3$ (or $\sigma^2$),
playing the role of the metric on $\DC^{1.1}$,
the spinor indeces $\alpha,\beta$ are raised and lowered by use of this
 metric.

On this space, we define the linear symplectic action of the
(2+1)-dimensional Poincare' group
\begin{equation}
\{P^a, P^b\}=0,\quad \{P^a, J^b\}=\varepsilon^{abc}P_c,
\{J^a, J^b\}=\varepsilon^{abc}J_c ,\label{2+1}\end{equation}
where
\begin{equation}
P^a=\pi^\alpha\sigma^a_{\alpha\bar\beta}{\bar\pi}^{\bar\beta}
\quad J^{a}=\frac{1}{2}\sigma^a_{\alpha\bar \beta}
(\pi^\alpha  {\bar\omega}^{\bar\beta}+ {\omega}^\alpha {\bar\pi}^{\bar\beta})
\label{5},
\end{equation}
and vector indices $a,b,c =0,1,2$, (or  $a,b,c=0,1,3$ at $\eta=\sigma^2$)
are raised and lowered by use of the metric
$g_{ab}=\frac 12 tr(\eta\sigma^a\eta\sigma^b)$,
where $\sigma^0_{\alpha\bar\beta}=\delta_{\alpha\bar\beta}$.

It can be verified easily that the invariants of the Poincare'group obey the
equalities $ P^aP_a=P^2,\;\; P_aJ^a=PJ$, where
\begin{eqnarray}
&P=\pi^\alpha\eta_{\alpha{\bar\beta}}{\bar\pi}^{\bar\beta},\;
\;J=\frac{1}{2}\eta_{\alpha{\bar\beta}}(\pi^\alpha{\bar\omega}^{\bar\beta}+
\omega^\alpha{\bar\pi}^{\bar\beta}), &\label{sm}\\
& \{P,J\}=0,\quad\{P, P^a\}=\{P, J^a\}=0, \quad \{J, P^a\}=\{J, J^a\}=0.&
\label{full}\end{eqnarray}
Therefore functions $J$ and $P$ can be interpreted as generators of
the anyon spin and mass, whereas the constraints
$ P^aP_a=m^2(>0),\quad P^aJ_a=ms$ fixing the anyon mass $m$ and
spin $s$ ($c=\hbar=1$) can be replaced by the equivalent ones
\begin{equation}
J= s, P= m,\;\;{\rm or} \;\; J=-s, P=-m, \label{ae}\end{equation}
which generate a linear transformations of the twistor space.

Note that the constraints reduce the number of (phase) degrees of
freedom down to four, and consequently, completely define the dynamics
of free anyon. The first pair of equations
(\ref{ae}) changes into the second pair via the discrete
canonical transformation corresponding to spatial reflections.
Thus we restrict ourselves by  the consideration of the first pair
of constraints in (\ref{ae}), putting $m>0$.

We will call below the above-described model of anyon on the twistor space
the $T-$model. \\

Let us perform the Hamiltonian reduction of the $T-$model by the action
of the spin generator $J$.

Since the generators
\begin{equation}
X^{a}=\frac{i}{2}\sigma^a_{\alpha{\bar\beta}}(\pi^\alpha{\bar\omega}^{\bar\beta}-
\omega^\alpha{\bar\pi}^{\bar\beta})
\end{equation}
obey the condition $\{X^a, J\}=0$, we take coordinates of the
reduced (six-dimensional) space to be the functions
$P^a, Q^a=\frac{X^a}{2P}$.

As a result, upon reduction of the Poisson brackets and generators $J^a$
we obtain the $K-$model described in the Introduction.

Now consider the quantum-mechanical $T$-model of the anyon in
the momentum representation
\begin{equation}
 {\bar\omega}_{\alpha} =-\frac{\partial}{\partial\pi^\alpha},
\quad {\omega}^{\alpha} =\frac{\partial}{\partial{\bar\pi}_{\alpha}},
\label{mr}\end{equation}
and impose, on the wave function $\Psi(\pi,{\bar\pi})$,
the analogs of constraints (\ref{ae}):
  \begin{equation}
P\Psi(\pi,{\bar\pi})= m\Psi(\pi,{\bar\pi}),
\quad {J}\Psi(\pi,{\bar\pi}) = s\Psi(\pi,{\bar\pi}).
\label{qr}\end{equation}

For reduction to the $K$-model we solve the second equation of (\ref{qr})
by the substitution
$$\Psi_s(p,{\bar p})= \psi_s(P^a){\rm e}^{is\gamma}, \quad [J,\gamma]=i.$$
Thus $\psi_s(P^a)$ represents the wave function  of the anyon
in the $K-$model, and the coordinates $Q^a$ are quantized as follows:
$${\hat Q}^a=-i\frac{\partial}{\partial P^a}-sA_a(P),
\quad A_a(P)=\{Q_a,\gamma\}.$$
Choosing, for example, $\gamma=\log{\pi^0}/{\bar\pi}^0$,
we shall see, that $A_a(P)$  is similar to the potential of
Dirac  monopole, and has the unique singularity in $P^0=P$
point.

Note that the generators $P^i=(P, P^a)$,
$J^{ij}=-J^{ji}=(J^{0a}=X^a, J^{ab}=\frac 12\epsilon^{abc}J_c)$
satisfy the relations $P^iP_i=0$ , $\{J, P^i\}=\{J, M^{ij}\}=0$
and define the algebra
$$\{P^i,P^j\}=0,\quad \{P^k, J^{ij}\}=2g^{k[i}P^{j]}, \quad
 \{J^{ij}, J^{kl}\}=2g^{k[i}M^{j]l}-2g^{i[k}M^{l]i},$$
where $ g_{ij}= diag(g_{ab}, -1)$.                                                               .

Thus at  $\eta=E$, $a=1,2,3$ they give the well-known twistor
representation of the (3+1)-dimensional Poincare' algebra for
massless particles, and $J$ represents the helicity operator. This
explains the repeatedly mentioned resemblance between the anyon covariant
model and the free model of a 3-dimensional massless particle. \\

{\bf{3.${\cal L}-$model.}} Solution of the mass constraint in the $K-$model
produces a noncovariant model of the anyon with the cotangent bundle
of the Lobachevsky plane as a free (without constraints)
phase space. However, it is more convenient to
derive it directly from the $T-$model through its Hamiltonian reduction
by  the action of generators $P$ and $J$ with values (\ref{ae}).
Owing to the commutativity of these generators, the reduced space is
four-dimensional, and consequently, it is a physical phase space of the anyon.

The complex coordinates of the reduced space $(p, z^*)$ can be take to be
the  functions
\begin{equation}
 p=\pi^1/\pi^0,\quad   z^*=\{p,\omega^\alpha{\bar\omega}_\alpha\}=
{\lambda}^1-\omega{\lambda}^{0},\quad\lambda^\alpha=\omega^{\alpha}/\pi^0,
\end{equation}
so that the reduced space turns out to be the tangent
bundle $T{\cal L}$ of the Lobachevsky plane ${\cal L}=SU(1.1)/U(1)$
in the Poincare' model
(or in the Klein model at $\eta=\sigma^2$) (see Appendix).
We identify $T\DC^{1.1}$ with cotangent bundle
$T_*{\cal L}$ by passing to the coordinates $p, z=ig(p,\bar p){\bar z}^*$,
where $g(p,\bar p)dpd{\bar p}$ is the metric of the Lobachevsky plane.

In these coordinates the symplectic structure complying with the
 reduced Poisson brackets is of the form
\begin{equation}
\Omega^{\rm red}= dz\wedge dp+d{\bar z}\wedge d{\bar p}+
i\frac sm g(z,{\bar z})dz\wedge d{\bar z},\label{rb}.\end{equation}

The generators  $P^a$ are reduces to the generators of the
 $SU(1.1)$ action
on ${\cal L}$ (see Appendix A), and  $J^a$ are reduced to the form
\begin{equation}
J^a=V^a(p)z+{\bar V}^a({\bar p}){\bar z}+ \frac{s}{m}P_a,
\label{pr}\end{equation}
where  $V^a(p)=i\partial_p P^a(p,{\bar p})$ define the isometries of the
Ka\"hler structure on ${\cal L}$.

So, the physical phase space of a free massive  relativistic  anyon
 is the cotangent bundle of the Lobachevsky plane with a
twisted symplectic structure.

 Quantum $T-$model (\ref{mr}) can be reduced
to the quantum ${\cal L}-$model, {\it e.g. } by the substitution
\begin{equation}
\Psi (\pi,{\bar\pi})=\psi_{m,s} (p, {\bar p}) e^{is\gamma},
\label{wfs}\end{equation}
which immediately results in
\begin{equation}
{\hat z}=i\frac{\partial}{\partial p}+
\frac{s}{m}\frac{\partial K(p,\bar p)}{\partial p},
\end{equation}
where $K(p,{\bar p})$ is the K\"ahler potential of the
Lobachevsky plane.

THe role of  wave function in such a quantum ${\cal L}-$model plays
 $\psi(p,\bar p)$, and  the generators $J^a$ define
 the (reducible) representation of the $(2+1)-$dimensional
Lorentz group on the Lobachevsky plane.

Note is to be made that this representation has arisen
in the extended models of anyons from quantization of
the isotopic space, where the parameter $p$ played the role of
isotopic coordinate. \\

{\bf 4.Conclusion.} So, we have presented the twistor
model for a free relativistic anyon
and have established correspondence between that model and
the minimal covariant model of anyon.
 We have found that the physical phase space of a classical free anyon
is a cotangent  bundle of the Lobachevsky plane with a twisted
symplectic (pseudo-K\"ahler) structure (the ${\cal L}-$model),
and its quantum mechanics is described by irreducible representations
of the (2+1)-dimensional Poincare' group.

Just as a $(3+1)-$dimensional massless particle is dual to the
system "charge-Dirac monopole", a massive relativistic anyon is dual to a
nonrelativistic particle moving on the Lobachevsky plane in a presence of
constant  uniform  magnetic field.

The $T-$model can be easily generalized for the anyon on
the (anti-)de-Sitter space. For this purpose note, that the generators
$J^a, P^a_{\pm}=\sigma^a_{\alpha{\bar\beta}}(\pi^\alpha{\bar\pi}^{\bar\beta}
\pm\omega^\alpha{\bar\omega}^{\bar\beta}),$ realize the algebras
$so(2.2)$ and $so(1.3)$ on the twistor space ${\bf T}$. In analogy with
the $T-$model, one can replace the  invariants
$P^a_{\pm} P_{a\;{\pm}}-J^aJ^a, P^a_{\pm}J_a$  by
$J, P_{\pm}=
\pi^\alpha{\bar\pi}_{\alpha} \pm\omega^\alpha{\bar\omega}_{\alpha}$.
The phase space of the analog of the $K-$model is again $T_*\DR^{1.2}$
with symplectic structure (\ref{Rs}). However, the  momentum
and mass generators are determined   in analogy with the Runge-Lenz vector  and the Hamiltonian
in the  "charge -dyon" (MIC-Kepler)  system\cite{nt},
but not by the  $P^a, P$ generators;
the physical phase space  is   ${\cal L}\times{\cal L}$.

 Note that  in the framework of the $T-$model,
on can easily to formulate the problem of interaction of a
relativistic anyon with an external uniform electromagnetic field,
 which became a subject of discussion \cite{ch},
but was not solved compleetely even at   the classical level.
\\

{\large Acknowledgements.} The author is grateful to A.Karabegov,
S.Lyakhovich and  A.Polychronakos for useful discussions
 as well as to E.Ivanov, A.Pashnev, V.Ter-Antonyan, and O.Khudaverdian
for interest to the work.  \\

 \vspace{0.5cm}
{{\bf Appendix. Lobachevsky plane as a reduced space.}}\\

 Consider the space $\DC^{1.1}$ with the pseudo-K\"ahler 2-form
 $\Omega^0= id\pi^\alpha\wedge d {\bar\pi}_\alpha$,
 where $ {\bar\pi}_\alpha=\eta_{\alpha\beta}{\bar\pi}^{\bar\beta}$,
and the metric $\eta$ of this space is the Pauli matrice.

The symplectic action of group $U(1.1)$ on this space
is given by the Hamiltonians
\begin{equation}
P=\pi^\alpha{\bar\pi}_\alpha,\;\;
 P^a=\pi^\alpha (T^a)^\beta\alpha {\bar\pi}_\beta:\quad
\Omega_0^{-1}(dP, dP^a)=0,\;\;
 \Omega_0^{-1}(dP^a, dP^b)=\varepsilon^{abc}P_c,
\label{p}\end{equation}
where $T^a= \sigma^a\eta$, $\sigma^a=( \sigma^0,...\neq {\eta},...3)$,
and indices are raised and lowered by use of the metric
$g_{ab}=tr{T^a T^b}=diag(+1,-1,-1)$.

The Hamiltonian  reduction of  $\DC^{1.1}$ by  the action  of
$U(1)$ given by the Hamiltonian $P$ corresponds to the factorization of
the level surface  $P=m$ by  the $U(1)$ group action.
Thus the reduced space is two-dimensional, and the (complex) functions
 \begin{equation} p=\frac{\pi^1}{\pi^0}\quad, {\bar
 p}^+=\frac{\bar\pi_{1}}{{\bar\pi}_{0}}
 \end{equation}
 can play the role of its local coordinates (when $m\neq 0$).
The action of the group $SU(1.1)$ on the reduced space is
given by linear-fractional transformations.

The reduced Poisson bracket is determined by the relations
 \begin{equation}
\{p,{\bar p}^+\}_0^{\rm red}=
\{p(\pi),{\bar p}^+(\bar\pi)\}_0\mid_{P=m}=
-\frac{i}{m} (1+p{\bar p}^+)^2,
\end{equation}
so that
$$g(p,{\bar p})dp d{\bar p}^+={m}/{(1+p{\bar p}^+)^2}dp d{\bar p}^+$$
defines the K\"ahler structure with the potential $K=m\log(1+pp^+)$
on the reduced space.

The symplectic action of the group $SU(1.1)$  on the reduced space
is generated by generators in $P^a$, restricted
to the level surface $P=m$.

Let $\eta=\sigma^3$. Then ${\bar p}^+=-{\bar p}$, and $|p| <1$ $(|p|>1)$
for  $m>0$ $(m<0)$, i.e. the reduced space is the Poincare'model of
the Lobachevsky plane.

The generators  $P^a$ on this space are of the form
\begin{equation}
P^0=m\frac{1+p{\bar p}}{1-p{\bar p}},\quad
P^1=m\frac{p+{\bar p}}{1-p{\bar p}},
\quad P^2=m\frac{i(p-{\bar p})}{1-p{\bar p}},
\end{equation}
and generates the isometries of K\"ahler structure
$\Omega_0^{-1}(dP^a,\;\; )
=V^a(p){\partial}_p+ {\bar V}^a({\bar p})\partial_{\bar p}$

Let $\eta=\sigma^1$, then ${\bar p}^+=-\frac{1}{\bar p}$,
and for $m>0$ $(m<0)$ we have $({\rm Im} p>0)({\rm Im} p<0)$ :
the reduced space is the Klein model of the Lobachevsky plane.

Analogously, at $\eta=\sigma^1$ we obtain the Klein model
described by the right (left) half-plane of the complex plane.

Note that the mapping of the interior of the circle into its
exterior in the Poincare'model (the upper half-plane into the
lower one in the Klein model), $m\to -m, \omega\to\frac{1}{\omega}$,
is corresponds to the spatial reflections:
 $P^0\to P^0,\; P^1\to P^1,\; P^2\to -P^2$ .\\

 \end{document}